\documentclass[twocolumn,apj]{openjournal}
\usepackage{graphicx} % Required for inserting images

\usepackage{orcidlink}

\def\gax{\mathrel{\raise.3ex\hbox{$>$}\mkern-14mu\lower0.6ex\hbox{$\sim$}}}
\def\lax{\mathrel{\raise.3ex\hbox{$<$}\mkern-14mu\lower0.6ex\hbox{$\sim$}}}
\def\gtorder{\mathrel{\raise.3ex\hbox{$>$}\mkern-14mu
             \lower0.6ex\hbox{$\sim$}}}
\def\ltorder{\mathrel{\raise.3ex\hbox{$<$}\mkern-14mu
             \lower0.6ex\hbox{$\sim$}}}
             
\begin{document}

\title{The HST Ultraviolet Spectrum of V723 Mon: Additional Evidence of a Stellar Companion}

\author{
C. S. Kochanek$^{1,2}$\orcidlink{0000-0001-6017-2961},
K. Z. Stanek$^{1,2}$\orcidlink{0009-0001-1470-8400}
T. A. Thompson$^{1,2}$\orcidlink{0000-0003-2377-9574}
T. Jayasinghe$^3$\orcidlink{0000-0002-6244-477X}
}
\affiliation{
  $^{1}$Department of Astronomy, The Ohio State University, 140 W 18th Ave, Columbus, OH 43210 \\
  $^{2}$Center for Cosmology and AstroParticle Physics, 191 W Woodruff Ave, Columbus, OH 43210 \\
  $^{3}$Independent Researcher, San Jose, CA
  }

\begin{abstract}
V723~Mon is a high mass function ($f=1.7 M_\odot$) single lined spectroscopic binary with a red giant primary that \cite{Jayasinghe2021J} suggested had  a black hole as its massive companion.  Unfortunately, \cite{elBadry2022} demonstrated that it had a hotter stellar 
companion whose detectability in optical spectra was difficult
due to its rapid rotation.  Here we confirm the presence of the
stellar companion with a Hubble Space Telescope STIS ultraviolet
spectrum.
\end{abstract}

\maketitle
~              

\section{Introduction}

The study of stellar-mass black holes has advanced significantly since their theoretical prediction in the early 20th century and the first observational evidence in the 1960s. Black holes were initially proposed as the endpoint of massive stars that undergo gravitational collapse \citep{Oppenheimer1939}, with Cygnus X-1 becoming the first robust stellar-mass black hole candidate identified through X-ray observations and radial velocity measurements of its companion \citep{Bolton1972, Webster1972}. For decades, the known population of stellar-mass black holes was limited to X-ray binaries, the accretion fed by the companion star providing an illuminating way to find them. However, this very limited sample represents a strongly biased subset of the full population. More recently, interest has turned to non-interacting black holes in binaries—objects not actively accreting and thus electromagnetically quiet.

Theoretical models predict that the Milky Way should contain a significant population of stellar-mass black holes, with estimates ranging from $\sim10^7$ to $\sim10^8$ for their total number, depending on assumptions about the initial mass function, star formation history, and binary evolution \citep[e.g.,][]{Breivik2017}. The vast majority of these black holes are expected to be isolated and quiescent, but possibly as many as 10\% of them might reside in binaries. Recent gravitational-wave detections by LIGO/Virgo have confirmed that massive stellar remnants can form in binary systems that eventually merge \citep{abbott2016, abbott2023}, supporting the idea that many binaries containing black holes remain undetected in the Galaxy. However, even with gravitational wave data our observational census remains biased and incomplete, and identifying nearby, quiescent black holes remains one of the key challenges in modern astrophysics. 

There are now a trickle of discoveries of non-interacting black hole binaries. The first were the discoveries of likely
systems in globular clusters by \cite{Giesers2018} and \cite{Giesers2019}.  This was followed by the discovery (\citealt{elBadry2023}, \citealt{elBadry2023}) of systems using binaries selected from the Gaia catalog of astrometric binaries (\citealt{Halbwachs2023}). 
There is also a good candidate for an isolated black hole identified through gravitational microlensing (\citealt{Lam2022}, \citealt{Sahu2022}).
Detecting dormant black holes in binaries is notoriously difficult. Most candidate identifications rely on precise radial velocity curves of the visible stellar companion, whose motion may indicate the presence of an unseen massive object. However, such detections are prone to multiple sources of systematic error. For example, in single-line spectroscopic binaries a lower luminosity stellar companion, neutron star or white dwarf can mimic the radial velocity signal of a black hole.  Recent surveys have uncovered many such false positives—initially proclaimed as black holes, but later reclassified as systems with compact stellar remnants or unusual stellar configurations. That was perhaps to be expected, given how rare ($\lesssim 1/1000)$ stellar-mass black holes are compared to regular stars, but in some cases Nature seems both subtle and malicious.

\citet[hereafter J21]{Jayasinghe2021J} proposed that V723~Mon has a $3 M_\odot$ dark companion most easily explained as a black hole. V723~Mon is a nearby ($d\simeq460\,\rm pc$), bright ($V\simeq8.3$~mag), red giant ($T_* \simeq4400$~K, $L_*\simeq170~L_\odot$) in a high mass function, 
$f(M)=1.72\pm 0.01~M_\odot$, nearly circular binary ($P=59.9$~d, $e\simeq 0$). J21 estimated that the companion has a mass of $M_c=2.91\pm 0.09~M_\odot$, and therefore was potentially a mass-gap black hole. However, subsequent work by \cite{elBadry2022} challenged this interpretation. They demonstrated that the observed data can be reconciled with a hotter stellar companion, eliminating the need for a black hole companion.
In this paper we analyze an ultraviolet spectrum of V723~Mon obtained with the Space Telescope Imaging
Spectrograph (STIS) on the Hubble Space Telescope (HST).  The data and analysis are presented in \S2,
and a brief discussion is give in \S3.

\begin{figure*}
    \centering
    \includegraphics[width=1.0\textwidth]{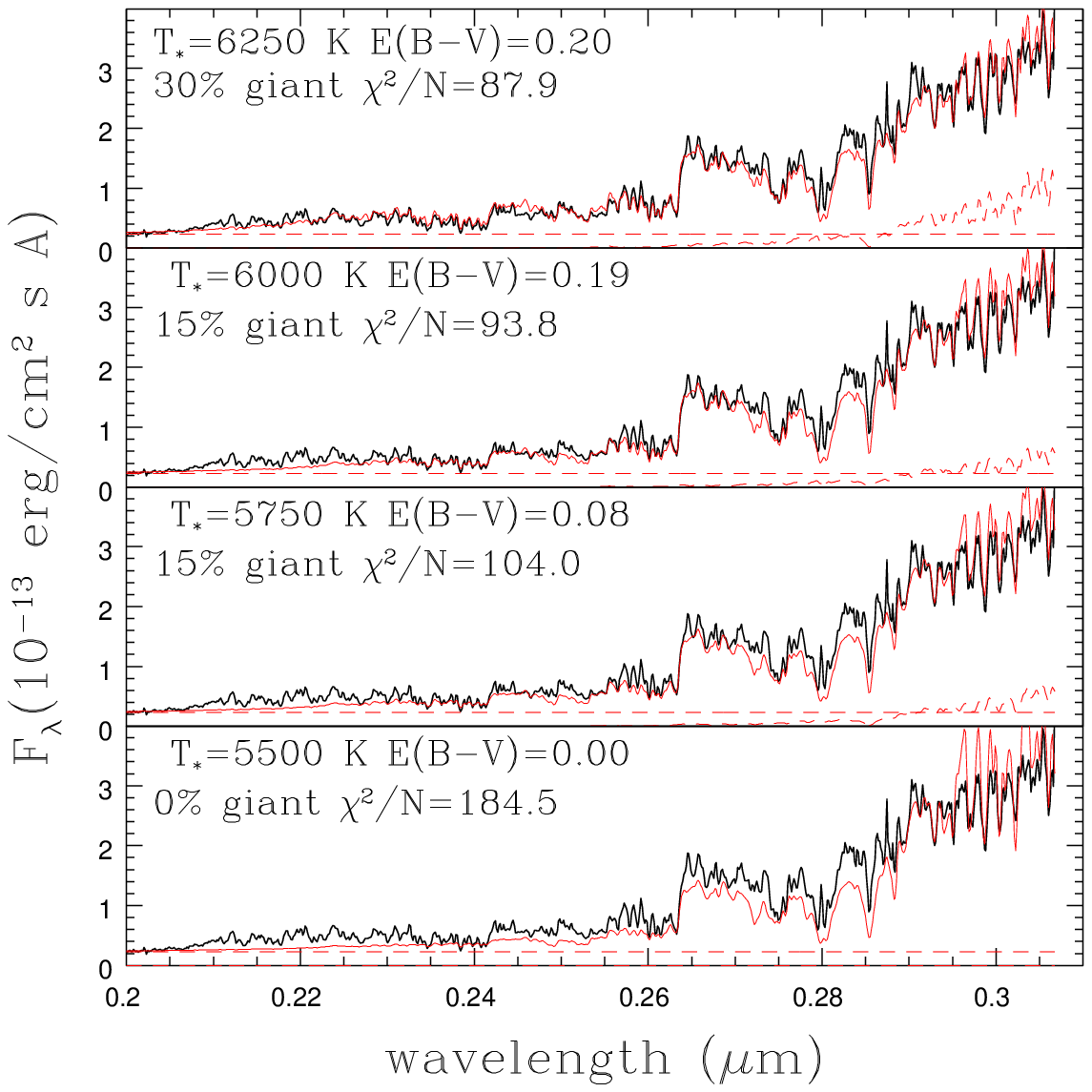}
    \caption{The STIS spectrum of V723~Mon (black) and the best models (red solid) for companion
        temperatures from $T_*=6250$~K (top) to $T_*=5500$~K (bottom). The dashed red lines
        show the contributions from the scattered light and the red giant.  The stellar temperature
        $T_*$, extinction $E(B-V)$, giant fraction relative to the companion ($\%$ giant), and the
        formal $\chi^2/N$ per degree of freedom are given in each panel.
          The spectrum extends to roughly $1600$\AA\ but is flat at the amplitude of the
        scattered light, so the shorter wavelengths are not shown in order to make the stellar emission more visible.
        }
    \label{fig:spec1}
\end{figure*}

\section{Data and Results}

We used the Space Telescope Imaging Spectrograph (STIS, \citealt{Woodgate1998}) CCD detector with the G230LB grating to obtain a
1685\AA\ to 3064\AA\ spectrum of V723~Mon using the
$52\times 0.2\hbox{E}1$ slit to minimize 
charge transfer efficiency (CTE) problems.  We used
651~second exposure times and the goal was to obtain
two spectra at each of three dither positions to control
cosmic rays and hot pixels.
Three spectra were sucessfully obtained on 10 December 2021 despite a failed guide star acquisition. 
Observations on 15 January 2022 also had a failed guide star acquisition, but in this case the STIS aperture door remained shut.  Repeat observations on 16 March 2022 sucessfully obtained the remaining three spectra.  

 We compared the 6 pipeline-reduced spectra to isolate 9 bad pixels distributed across the spectra and averaged the non-bad pixels. The source is red, so a UV spectrum with the G230LB grating will contain a significant amount of scattered light, leading to an upturn at the shortest wavelengths plus a roughly constant contribution at all wavelengths (see \citealt{Heap2016}). We modeled and subtracted the upturn, and include a constant term when fitting the spectra. We found no detectable wavelength shift between the two epochs of spectra, which is not surprising given that the STIS spectral resolution is
$R\sim 900$ ($\sim 330$~km/s) while \cite{elBadry2022} argue that $K_2 \ltorder 11$~km/s. The resulting mean spectrum after removing the upturn is shown in Fig.~\ref{fig:spec1}.

Based on the parameters found by \cite{elBadry2022}, we extracted the $\log(g) = 3$, $[\hbox{Fe/H}]=-0.5$ metallicity model stellar atmospheres from 
\cite{Prieto2018}.  We removed a $2.3$\AA\ wavelength offset (determined by cross correlating the data and the models) between the observed spectrum and the models, smoothed
the higher spectral resolution models with a Gaussian of dispersion $\sigma$ and allowed for extinction $E(B-V)$  using a $R_V=3.1$ \cite{Cardelli1989} extinction curve.  We fit the spectrum as a function of the stellar temperature $T_*$ ($5000$~K to $6750$~K in steps of $250$~K), extinction ($0.00 \leq E(B-V) \leq 0.20$), smoothing ($1.5\hbox{\AA} \leq \sigma \leq 2.5\hbox{AA}$) and a constant for the scattered light.  The latter parameter was restricted to lie between zero and the level of the plateau seen at the shorter wavelengths.  In all the models we also included a $T_*=4000$~K component for contamination from the giant with its unextincted flux ratio relative the companion over the wavelength range from $2950$\AA\ to $3050$\AA\ ranging from $r=0$ to 30\% in steps of 5\%. The fraction of the flux from the giant is then $f=r/(1+r)$.  Larger extinctions or giant contributions are ruled out by the results of \cite{elBadry2022} and our SED models below.

Fig.~\ref{fig:spec1} illustrates the basic results using the best models with temperatures from $T_*=5500$~K to $6250$~K.    While there is no perfect match, the features of the model spectra closely match the features seen in the data. Of the four examples shown in Fig.~\ref{fig:spec1}, the coldest $T_*=5500$~K model fits the worst, and the problems worsen for still colder temperatures.  The
best $T_*=5500$~K model has no extinction and no contribution from the giant, but still significantly underpredicts the UV flux. We also know from \cite{elBadry2022} and our own models below, that the overall spectral energy distribution requires both finite extinction and a contribution from the giant towards the red end of this wavelength range.

The model with $T_*=5750$~K is close the temperature estimate in \cite{elBadry2022} and has similar extinctions and giant fractions ($r=15\%$, so $f=13\%$). However, the model still has trouble matching the
bluer fluxes of the data.  The still hotter models have intrinsic spectra which are flatter, so the models must increase the amount of giant contamination and the extinction to match the observed shape.
The hottest, $T_*=6250$~K model arguably matches the bluest parts of the spectrum best, but the extinction
is now high compared to the SED models (in \cite{elBadry2022} and below), and with $r=30\%$, the giant flux fraction of $f=23\%$ is getting too high.

None of the models
are perfect fits to the spectra given their uncertainties, with $\chi^2/N \sim 100$.  A common problem for all the models is that none match the shortest wavelengths (around 2200\AA) very well. 
This is near the $2175$\AA\ extinction feature, so we tried varying $R_V$ to see if the extinction model was driving the problem, but it made little difference. We also tried Solar metallicity models.  These fit the spectra modestly better, but had the same basic patterns seen in Fig.~\ref{fig:spec1} for the goodness of fit, variations in the extinction and the giant contamination, and the difficulties fitting the shortest wavelengths.  Nonetheless, it is clear that the UV spectrum of V723~Mon is dominated by by a star with a temperature of 
$T_* \simeq 5750$~K to $6000$~K and so require a hot, stellar companion to the giant primary, as concluded by \cite{elBadry2022}.

\begin{table}[t]
  \centering
  \caption{Photometry of V723~Mon}
  \begin{tabular}{lc}
  \hline
  \multicolumn{1}{c}{Band} &
  \multicolumn{1}{c}{Magnitude} \\
  \hline
  AllWISE F22W &$4.823\pm0.035$ \\
  AllWISE F12W &$5.008\pm0.015$ \\
  AllWISE F45W &$5.096\pm0.072$ \\
  AllWISE F34W &$5.284\pm0.157$ \\
  2MASS K$_s$  &$5.364\pm0.021$ \\
  2MASS H      &$5.576\pm0.034$ \\
  2MASS J      &$6.259\pm0.027$ \\
  APASS z      &$7.360\pm0.051$ \\
  APASS i      &$7.488\pm0.138$ \\
  APASS r      &$7.886\pm0.117$ \\
  APASS V      &$8.301\pm0.035$ \\
  APASS g      &$8.730\pm0.007$ \\
  APASS B      &$9.244\pm0.044$ \\
  SkyMapper v  &$10.000\pm0.019$ \\
  SkyMapper u  &$10.456\pm0.040$ \\
  Swift UVM2   &$14.11 \pm 0.07$\\
  STIS F275W   &$12.74$ \\
  STIS F225W   &$14.30$ \\
  \hline
  \multicolumn{2}{l}{The griz, Swift and STIS magnitudes} \\
  \multicolumn{2}{l}{are on the AB magnitude system,}\\
  \multicolumn{2}{l}{and the rest are on the Vega system } \\
  \end{tabular}
  \label{tab:phot}
\end{table}

\begin{figure}
    \centering
    \includegraphics[width=0.45\textwidth]{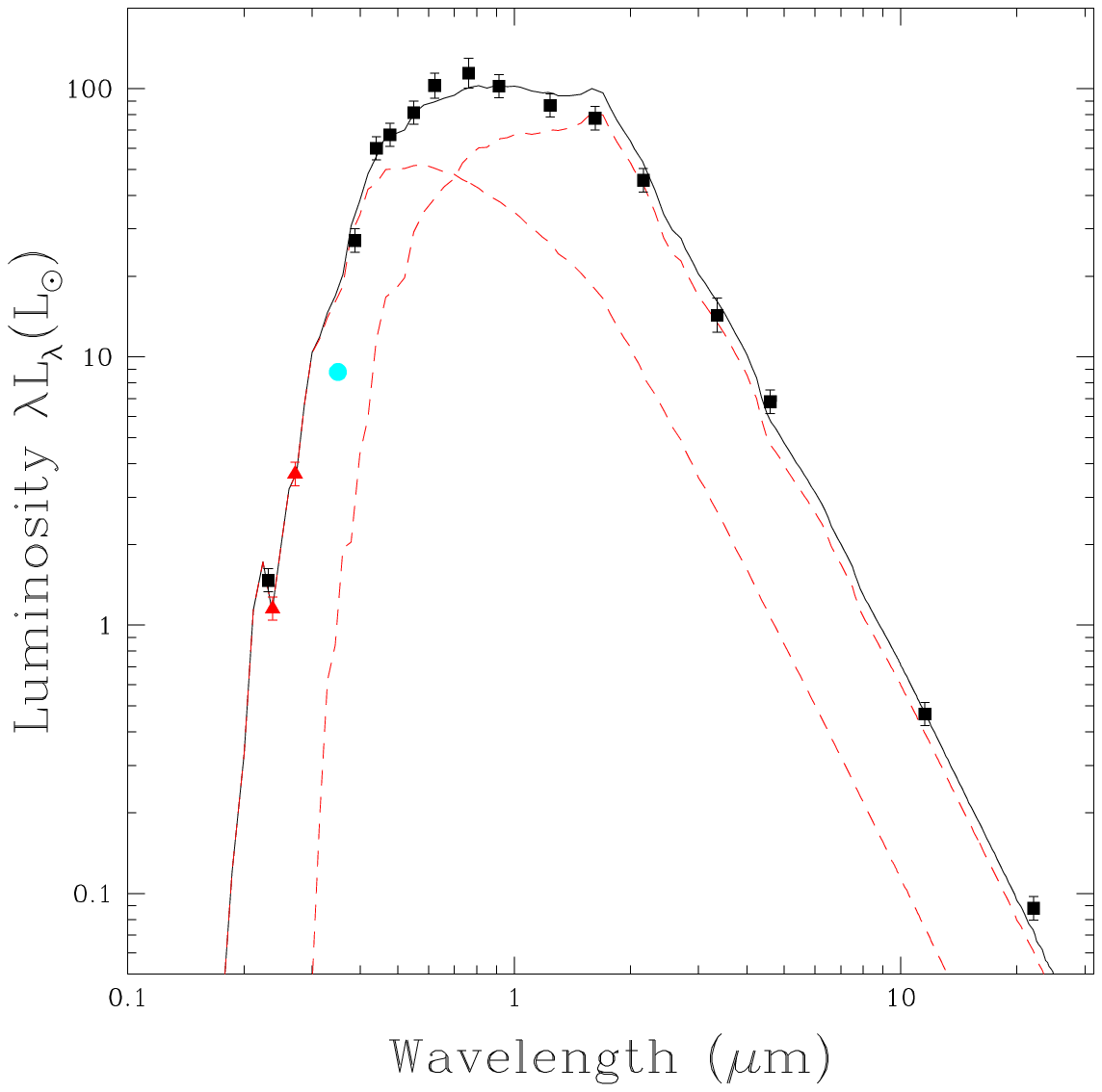}
    \caption{The SED of V723~Mon (black solid line) and the SEDs of the two component
        stars (red dashed lines) as compared to the synthetic HST luminosities (red
        triangles) and the other photometric data (black squares).  The cyan circle
        is the SkyMapper u band point that was dropped from the fits. The SED is
        corrected for the model extinction.
        }
    \label{fig:sed}
\end{figure}

Next we modeled the spectral energy distribution (SED) including the HST results.
Table~\ref{tab:phot} provides the photometry used for the SED model.  We use the AllWISE (\citealt{Wright2010}) mid-IR magnitudes, the 2MASS (\citealt{Skrutskie2006}) near-IR magnitudes, the APASS DR10 (\citealt{Henden2018}) optical magnitudes, the SkyMapper DR4 (\citealt{Onken2024}) $u$/$v$ band magnitudes, and the UVM2 Swift magnitude from \cite{Jayasinghe2021J}. We synthesized HST F225W and F275W AB magnitudes from the STIS spectra. These have negligible statistical errors and are consistent with the earlier Swift UVM2 magnitude. \cite{elBadry2022} note that the SkyMapper data may be saturated. We ultimately drop the 
SkyMapper u band point as an outlier, but had no difficulty fitting the SkyMapper v band data.

Assuming the photogeometric distance of $455.6\pm0.7$~pc from \cite{BailerJones2021}, we simultaneously fit these data using two stars with temperatures and
luminosities of ($T_{1*}$, $L_{1*}$) and ($T_{2*}$, $L_{2*}$) plus a foreground extinction of $E(B-V)$
with a \cite{Cardelli1989} $R_V=3.1$ extinction law.  We used 
$[\hbox{Fe/H}]=-0.5$ stellar atmosphere models from \cite{Castelli2003}. We assumed minimum fractional luminosity uncertainties which are the larger of the reported uncertainties and 10\%.   We used Gaussian priors  for the temperatures of $T_{*1}=3800\pm 100$ for the cool primary and $T_{*2}=5800 \pm 200$~K for the hotter secondary based on the spectral models of \cite{elBadry2022}.
We also used a Gaussian extinction prior of $E(B-V) = 0.10 \pm 0.04$ based on the three-dimensional dust models of \cite{Green2019}. We fit the the data in terms of $\log(\nu L_\nu)$ for each band. The fits and their uncertainties were obtained with Markov Chain Monte Carlo (MCMC) methods.

Fig.~\ref{fig:sed} shows the resulting best fit for the SED, and the contributions from the two
stars.  The two temperatures are $T_{*1}=3860 \pm 130$~K and $T_{*2}=5910\pm160$~K at 90\% 
confidence, with luminosities of $L_{*1}=10^{2.01\pm 0.07}$ and $L_{*2}=10^{1.83\pm 0.15}L_\odot$.
The extinction is $E(B-V)=0.10$ with a 90\% confidence range of $0.03$ to $0.16$. The temperatures
and the extinction are strongly correlated in the sense that the two temperatures tend to 
increase or decrease together while also increasing or decreasing the extinction.  While the temperature
posteriors are narrower than their priors, the exinction posterior is broader.

\section{Discussion}

The STIS ultraviolet spectrum of V723~Mon clearly shows the presence of the hot companion to the red giant 
discovered by \cite{elBadry2022} with very little contamination from the giant.  The spectroscopic fits are
imperfect, particularly near 2200\AA.  Attempts to solve this by varying the extinction law, and hence the
strength of the 2175\AA\ extinction, feature were not successful.  In fact, the UV spectrum of V723~Mon is very similar to that found by \cite{Bianchi2024} for the proposed non-interacting black hole binary 2MASS~J0521$+$4359 (\citealt{Thompson2019}).  The orbital velocity
of the companion can be determined with sparse higher spectral resolution STIS observations (likely two observations at quadrature with the G230MB grating and its $R\sim 6000$, instead of $R\sim 900$ here, 
resolution).  At least while HST is available, obtaining ultraviolet spectra should become a routine part of searches for non-interacting compact object binaries.

\section*{Acknowledgements}

 CSK and KZS are supported by NSF grants AST-2307385 and 2407206. This research is based on observations made with the NASA/ESA Hubble Space Telescope obtained from the Space Telescope Science Institute, which is operated by the Association of Universities for Research in Astronomy, Inc., under NASA contract NAS 5–26555. These observations are associated with program GO-116708.

\bibliography{main}{}
\bibliographystyle{aasjournal}

\end{document}